# A Dynamic Framework for Semantic Grouping of Common Data Elements (CDE) Using Embeddings and Clustering

Madan Krishnamurthy[1], Daniel Korn[1], Melissa A Haendel[1], Christopher J Mungall[2], Anne E Thessen[1]

[1]University of North Carolina at Chapel Hill, Chapel Hill, NC, USA

[2]Lawrence Berkeley National Laboratory, Berkeley, CA, USA

# Abstract

**Objective**: This research aims to develop a dynamic and scalable framework to facilitate the harmonization of Common Data Elements (CDEs) across heterogeneous biomedical datasets. By addressing challenges such as semantic heterogeneity, structural variability, and context-dependence, we seek to streamline the integration of diverse data sources to enhance interoperability and accelerate scientific discovery.

**Methods**: Our methodology leverages Large Language Models (LLMs) for context-aware text embeddings, which convert CDEs into dense vectors that capture semantic relationships and patterns. These embeddings are then used in unsupervised clustering to group semantically similar CDEs. The framework incorporates: (1) text embedding of CDE using LLMs to mathematically represent semantic context, (2) unsupervised clustering of these embeddings with Hierarchical Density-Based Spatial Clustering of Applications with Noise (HDBSCAN) to group similar CDEs, (3) automated labeling using LLM-based summarization, and (4) supervised learning to train a classifier that assigns new or unclustered CDEs to one of the labeled clusters. This approach facilitates CDE harmonization while minimizing manual effort.

**Results**: The framework was evaluated using the National Institutes of Health National Library of Medicine (NIH NLM) CDE Repository containing over 24,000 CDEs. To suggest CDEs for harmonization, semantically similar CDEs were grouped using unsupervised clustering, enabling automated identification of consistent and reusable data patterns across datasets without human annotation. With an optimized minimum cluster size of 20, the system identified 118 meaningful clusters. The classification model achieved an overall accuracy of 90.46%, performing exceptionally well in categories with larger sample sizes. External validation against Gravity Projects' Social Determinants of Health domains showed strong alignment with Adjusted Rand Index (ARI) of 0.52 and Normalized Mutual Information (NMI) of 0.78. The high classification accuracy achieved in this study demonstrates that the generated embeddings effectively capture cluster-specific characteristics.

**Conclusion**: This approach provides an adaptable solution to the ongoing challenge of CDE harmonization, enabling more efficient selection of CDEs for harmonization. The framework's scalability ensures it can accommodate future data growth, making it a valuable tool for enhancing data interoperability both prospectively and retrospectively. Future work should focus on addressing data imbalance and improving performance for underrepresented categories to further enhance the framework's utility across diverse biomedical domains.

# Introduction

The harmonization of data across diverse and heterogeneous sources remains a fundamental challenge in modern biomedical research. As biomedical datasets grow in volume and complexity, the need for standardized Common Data Elements (CDEs)[1] becomes ever more critical. CDEs are vital data descriptors generally used to standardize medical and social survey questions. These CDEs typically consist of several Data Elements, which serve as the building blocks of CDEs. The formalized format of these descriptors ensure interoperability across disparate datasets, promoting seamless data sharing and integration across various research studies, institutions, and repositories.

According to established metadata standards such as ISO/IEC 11179[2], a Data Element consists of both conceptual and representational components. At its core is the *Data Element Concept* (*DEC*), which captures the meaning of the data independent of any particular format. A DEC is further composed of an *Object Class* (e.g., a "Person") and a *Property* (e.g., "Smoking status"). The representational layer includes the *Value Domain*, which defines the allowable range of values, and associated *Permissible Values*, which are the specific accepted entries (e.g., "Current," "Former," "Never"). Additional specifications—such as *Data Type, Unit of Measure*, and *Representation Class*—ensure consistent interpretation and application of data. These structured components facilitate semantic clarity, but differences in terminology, categorization, and encoding across datasets—for example, representing smoking status as "Yes/No" versus "Current/Former/Never"—illustrate the persistent challenges to achieving true interoperability.

While CDE repositories such as National Institutes of Health National Library of Medicine (NIH NLM)[4], Cancer Data Standards Registry and Repository (caDSR)[5], Center for Expanded Data Annotation and Retrieval (CEDAR)[6], PHENotypes and eXposures (PhenX)[7], and Metadata Online Registry (METEOR)[8] offer standardized frameworks for data element definitions and metadata, the practical challenge of aligning heterogeneous data elements across these platforms is still largely unresolved[9]. This challenge is further amplified by several factors:

1. **Semantic Heterogeneity**: The same concept may be expressed using different terminologies or vocabularies across datasets. For example, "Age at diagnosis" could be described as "Patient's age at initial diagnosis" in one dataset and "Age at onset" in another.
2. **Structural Variability**: Equivalent data elements may appear in different formats, schemas, or organizational structures. A CDE in one dataset might be a free-text field, while in another, the same concept could be a numeric or categorical field.

3. **Context-Dependence**: The interpretation and significance of data elements may vary depending on the specific research domain or study context. For example, the term "subject" may refer to a patient in clinical trials but to a participant in a psychological study.

These challenges make them inadequate for large-scale and dynamic integration of CDEs. The need for scalable, automated solutions that can effectively address these complexities has never been more urgent, especially as contemporary biomedical datasets continue to grow in volume, diversity, and intricacy.

## Proposed Framework to Support CDE Harmonization

This paper introduces a dynamic and scalable framework to support CDE harmonization, leveraging Large Language Model (LLM) embeddings and unsupervised clustering techniques[10]. The framework clusters semantically similar CDEs and assigns contextual labels to guide their alignment—particularly useful in the absence of a unified target schema. From an AI-readiness perspective, the use of embeddings satisfies the requirement for numerical input in machine learning models, while the cluster names can function as proxy labels for downstream classification tasks. The proposed methodology incorporates the following steps:

1. **Text Embedding & Feature Representation**: Each CDE is transformed into a high-dimensional semantic vector using a pretrained LLM, which preserves its contextual meaning despite variations in naming conventions or formats.
2. **Unsupervised Learning (Clustering)**: To identify groups of semantically similar CDEs, unsupervised clustering is applied based on embedding distances. While our current implementation uses HDBSCAN[11] for its ability to handle clusters of varying density without requiring the number of clusters in advance, other algorithms such as K-means or DBSCAN[12] could be applied in different scenarios depending on the dataset characteristics.
3. **Automated Labeling**: Descriptive labels are generated for each cluster using LLM-based summarization techniques, significantly reducing the need for manual data annotation.
4. **Supervised Learning (Training Models)**: The labeled clusters are used as training data for predictive models (e.g., Random Forest, XGBoost, and Neural Networks)[13] to classify new CDEs.

Although the output does not constitute fully harmonized CDEs, it offers a structured, semantically-informed scaffold that supports both harmonization efforts and AI-driven analysis. This dynamic, end-to-end pipeline adapts to new data inputs, refining clustering as datasets evolve. Its scalability enables seamless integration of increasing CDE volumes and new repositories without major reconfiguration. Dimensionality reduction techniques like PCA and t-SNE[14] optimize computational efficiency while preserving semantic relationships for clustering and classification. By automating CDE transformation the framework minimizes manual curation effort, supports large-scale integration of previously unaligned data, and enhances cross-institutional collaboration. Leveraging semantic understanding and equivalences over precise overly rigorous syntactic matching, it improves data alignment, unlocking the value of siloed datasets and accelerating scientific discovery.

## Related work

Recent advancements in Large Language Models (LLMs)[15], such as OpenAI's GPT[15] models and transformers like BERT[15], offer promising solutions to these challenges by leveraging dense vector embeddings[16]. The transformation of textual descriptions into high-dimensional vector embeddings ensures that semantic relationships and contextual meaning are preserved despite terminological discrepancies. Despite the promising potential of LLM embeddings, their application to CDE harmonization remains underexplored, particularly in the development of end-to-end frameworks that integrate semantic embedding representations with automated processes for clustering, labeling, and classification.

A recent study has explored methodologies for improving semantic interoperability and harmonization of CDEs in biomedical contexts. CDEMapper proposed a framework for harmonizing biomedical datasets by leveraging embeddings to address challenges related to semantic heterogeneity and structural variability[17]. Other studies have highlighted the importance of semantic alignment for enhancing interoperability. For instance, the study "Toward Better Semantic Interoperability of Data Element Repositories in Medicine" analyzed the challenges in aligning CDE repositories and proposed methodologies for improving semantic consistency[9]. Similarly, "The roles of common data elements and harmonization"[1] explored the broader implications of CDE harmonization in clinical research, emphasizing the value of standardized CDEs for cross-institutional collaboration[1]. Machine learning techniques have also been applied to the task of CDE mapping. One study employed artificial neural networks to map cancer-related CDEs to the Biomedical Research Integrated Domain Group (BRIDG) model, offering a semi-automated approach to enhancing CDE interoperability[18]. Another investigation focused on Alzheimer's disease-related CDEs, mapping them to the NIH CDE framework and demonstrating the utility of automated techniques for aligning disease-specific datasets[19].

Furthermore, several repositories and initiatives have contributed to the development of standardized CDEs. Resources such as the NIA's Common Data Elements Webinar Series[20] and the ICPSR's CDE collection[21] offer platforms for understanding and adopting standardized CDE frameworks. Metadata registries, such as those discussed in "Metadata Registry" [22], provide structured environments for cataloging and accessing CDEs, further facilitating interoperability.

However, none of the aforementioned efforts provide guidance on which CDEs should be harmonized or implement the mechanisms necessary for achieving semantic interoperability across disparate datasets. Collectively, these related works highlight the increasing recognition of the importance of CDE harmonization and the need for scalable, automated approaches to manage the growing complexity of biomedical data. Building on these efforts, the proposed framework integrates LLM embeddings with unsupervised clustering techniques to offer a dynamic and scalable solution that addresses challenges such as semantic heterogeneity, structural variability, and context dependence.

# Methods

**Fig 1** outlines the process we utilized for getting CDEs from multiple repositories, transforming them into embeddings, clustering similar concepts, and assigning standardized labels. The steps involved are as shown in **Table 1.**

**Table 1.** Methodology steps

| Step | Input | Function | Output |
| --- | --- | --- | --- |
| Data Collection | CDEs are gathered from various repositories (e.g., NIH, PhenX). | The data is collected through API calls or data dumps and stored for further processing. | A dataset of CDEs in their raw form. |
| Embedding Transformation | Raw CDEs are passed into a pre-trained LLM (e.g., GPT, BERT). | The LLM is used to transform the CDEs into embeddings. | A set of embedding vectors that represent the CDEs, where semantically similar concepts are closer in the embedding space. |
| Clustering | The embeddings produced in the previous step. | A clustering algorithm (e.g., K-means, HDBSCAN) is applied to group similar CDEs based on their semantic similarity, derived from the embeddings. | Clusters of similar CDEs, where each cluster represents a group of related concepts. |
| Label Assignment | The clusters formed from the previous step. | Each cluster is assigned a standardized label, which could be a category or concept universally recognized in the biomedical domain (e.g., "Neurological Disorder" or "Educational Attainment"). | Standardized labels are assigned to each cluster, helping in the classification and further analysis. |

| Step | Input | Function | Output |
|------|-------|----------|--------|
| Validation | The clusters and their assigned labels. | A classification model (e.g., a supervised machine learning classifier) is used to validate the clusters and labels for consistency and accuracy, ensuring they align with domain-specific standards. | A validation score that confirms the quality and interoperability of the clusters and labels with other data models. |

Detailed methodology and technical implementation of these steps are provided in the respective subsections.

**Figure 1.** Workflow for embedding, clustering, and classifying CDEs across repositories..

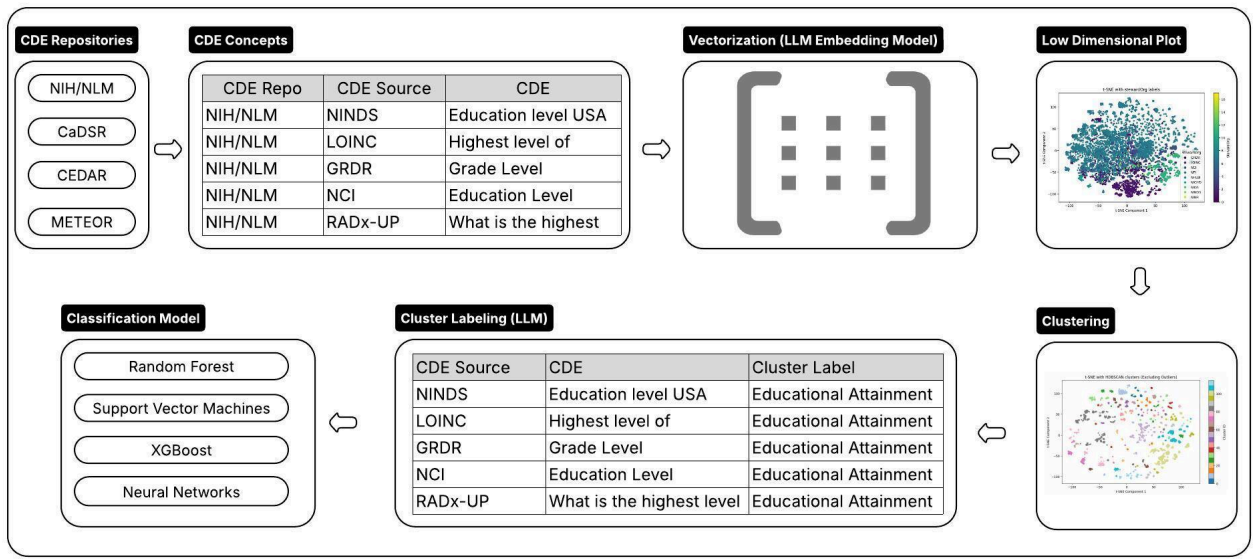

## Data Source: NIH NLM CDE Repository[23]

Common Data Elements (CDEs) from the NIH NLM CDE Repository were integrated into the framework to showcase its ability to process diverse datasets. The repository aggregates CDEs from various biomedical initiatives, offering standardized data across domains such as neurological disorders (NINDS)[24], heart and lung research (NHLBI)[25], rare diseases (GRDR)[26], and patient-reported outcomes (PROMIS/Neuro-QOL)[27,28], totaling 24,363 CDEs. Initially provided in JSON format, the data is transformed into a structured pandas DataFrame, extracting key elements to support downstream analyses.

Preprocessing

Data from the original JSON format was selectively processed to retain only essential elements: unique identifiers (*tinyId*), stewarding organizations (*stewardOrg*), *designations*, *definitions*, and *permissible values*. The *tinyId* serves as a compact, unique reference for each CDE, while *stewardOrg* identifies the organization responsible for maintaining the CDE. *Designations* refer to how a particular CDE is labeled. When multiple designations were available, the one tagged as "Preferred Question Text" was selected. If no such tag was present, the first available designation was used, ensuring clarity and relevance. A *definition* provides a precise and consistent description of the CDE, while *Permissible values* refer to the set of allowed responses or entries for a given CDE.

Additionally, permissible values were concatenated into strings for uniformity across the dataset. Each designation for a given CDE was represented as a distinct row in the final table, ensuring that the dataset was structured appropriately for downstream processing, including embedding generation.

Metadata fields not essential to the core interpretation of each CDE—including *property, objectClass, dataElementConcept, valueDomain.identifiers, valueDomain.ids, valueDomain.codeSystemName, classification, properties, sources, createdBy*, as well as provenance-related metadata such as *created, imported, views*, and *changeNote*—were intentionally excluded. While these components may provide additional context, their omission was aimed at reducing noise and optimizing the dataset.

**Table 2.** Sample of tabularly structured CDEs from NIH NLM CDE Repository to illustrate how the data is organized.

| tinyId | Designation | Definition | Permissible Values | Steward Org |
|---|---|---|---|---|
| 6AT_JFxD1 | State | The state for the address to describe where a mail piece is intended to be delivered. | AL: Alabama C43479, AK: Alaska C43506, AZ: Arizona C43505, etc. | Project 5 (COVID-19) |
| PDjBiGXjO | Age | The number of years or months (if 24 months or younger). | No permissible values | ScHARe |

## Embeddings Generation

Embeddings for the Common Data Elements (CDEs) were generated using OpenAI's `text-embedding-3-small` model[29]. This lightweight embedding model generates 1,536-dimensional dense vectors to represent the semantic meaning of text. The *designation*, *definition*, and *permissible values* fields were concatenated into a single string for each entry (joined with a whitespace between each value), which was then used as input for embedding generation. Missing values were replaced with empty strings, and all fields were cast to string format for consistency. The embeddings were generated through OpenAI's API. The resulting embeddings were stored in the *embedding* column for downstream tasks like clustering and similarity analysis.

## Clustering

Clustering was performed on the embeddings from the Common Data Elements (CDEs) to uncover patterns in the data using the Hierarchical Density-Based Spatial Clustering of Applications with Noise (HDBSCAN) algorithm[11]. HDBSCAN was chosen for its ability to identify clusters of varying densities and handle noise and outliers.

Unlike traditional methods like K-means, which assume spherical clusters and require the number of clusters to be predetermined, HDBSCAN can detect clusters of arbitrary shapes and densities without specifying the number of clusters in advance. It also automatically labels outliers with a value of -1, which is beneficial for noisy datasets. Outliers, represented as points that do not fit well into any cluster, can often signify anomalies or rare events. By explicitly labeling these points, HDBSCAN helps isolate noise from the meaningful structure of the data, enhancing the overall clustering process.

Similarity between CDE embeddings was computed using cosine distance. Clustering performance was evaluated using three internal validation metrics[30], which assess the quality and effectiveness of clusters. These metrics help determine how well the data is grouped into clusters based on its inherent characteristics:

- **Silhouette Score**[30]: Indicates how well-defined clusters are, with values close to +1 representing well-separated clusters.
- **Dunn Index**[30]: Measures inter-cluster separation relative to intra-cluster variance, with higher values indicating better separation.
- **Davies-Bouldin Index**[30]: Assesses cluster compactness, where lower values indicate better separation.

The *min_cluster_size* parameter plays a crucial role in controlling the granularity of the clusters. It specifies the minimum number of points required to form a valid cluster. This parameter was optimized by testing values from 5 to 500 (as shown in **Table 3**) to strike a balance between clustering quality, granularity, and outlier detection. By adjusting the *min_cluster_size*, we can influence the number of clusters formed and how many data points are treated as outliers. Larger values of *min_cluster_size* result in fewer, more general clusters, while smaller values increase the number of clusters, potentially capturing finer distinctions but also increasing the likelihood of outliers.

Best Balance Across All Metrics: *min_cluster_size* = 20 offers the best trade-off between cluster quality and quantity, with a high silhouette score (0.3064), a reasonable Dunn index (0.3335), and a relatively low Davies-Bouldin index (1.3714). This clustering configuration will form the foundation for the next phases of the analysis, guiding subsequent steps such as visualization, labeling, and classification. By leveraging this well-balanced setup, we ensure that the clusters remain both insightful and actionable, enhancing the overall analysis process.

**Table 3.** Internal Evaluation of Clustering Performance Across Different min_cluster_size Values

| min_cluster_size | Silhouette Score | Dunn Index | Davies-Bouldin Index | Number of Clusters | Number of Outliers |
| --- | --- | --- | --- | --- | --- |
| 5 | 0.2847 | 0.0007 | 1.3617 | 645 | 13251 |
| 10 | 0.2864 | 0.3390 | 1.3848 | 248 | 15567 |
| 15 | 0.2944 | 0.3671 | 1.4423 | 165 | 16824 |
| 20 | 0.3064 | 0.3335 | 1.3713 | 118 | 17973 |
| 25 | 0.2010 | 0.3884 | 1.5024 | 71 | 17430 |
| 50 | 0.1971 | 0.4696 | 1.5315 | 20 | 19731 |
| 75 | 0.1919 | 0.5296 | 1.6296 | 12 | 19795 |
| 100 | 0.2012 | 0.4507 | 1.9619 | 6 | 20506 |
| 250 | 0.1996 | 0.5814 | 1.5305 | 2 | 21099 |

This configuration also surfaced several noteworthy outliers—CDEs that did not meet the minimum cluster size of 20 and were labeled as noise by HDBSCAN. For example, the "Birth date" CDE in **Table 4** exhibits a highly specific definition that sets it apart from other elements, as there were not enough data points (20) to form a cluster. These cases are acknowledged as part of the clustering outcome and may reflect the distinct semantics of certain CDEs under the current configuration.

**Table 4.** Examples of outliers CDEs not falling under any cluster group. In this case it was because too few CDEs available in the NIH repository had semantic content related to birth dates.

| tinyId | Designation | Definition | Permissible Values | Steward Org |
|---|---|---|---|---|
| OtsN78xANu1 | Birth date | Date (and time, if applicable and known) the participant/subject was born | No permissible values | NINDS |
| Qyvxrsconjl | Birth date Family member | No definition available | No permissible values | LOINC |

Clustering Visualization

To explore the data before and after clustering, we applied dimensionality reduction techniques. Principal Component Analysis (PCA)[31] reduced the high-dimensional embeddings to 50 dimensions while preserving variance. Subsequently, t-Distributed Stochastic Neighbor Embedding (t-SNE)[31] projected these embeddings into two dimensions for visualization.

**Figure 2.** Shows the t-SNE plot before clustering, colored by *stewardOrg*, revealing organization-specific grouping patterns.

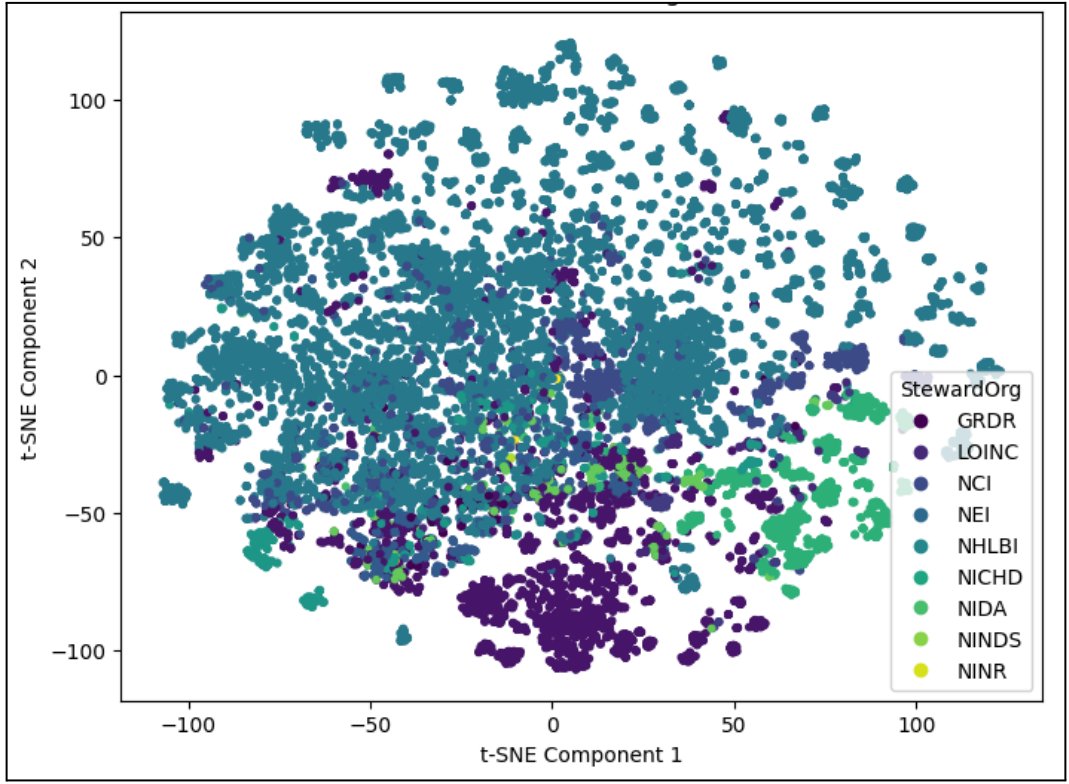

**Figure 3.** Visualizes the data after clustering with HDBSCAN, excluding outliers, and highlights distinct clusters (118 clusters), indicating meaningful groupings in the data.

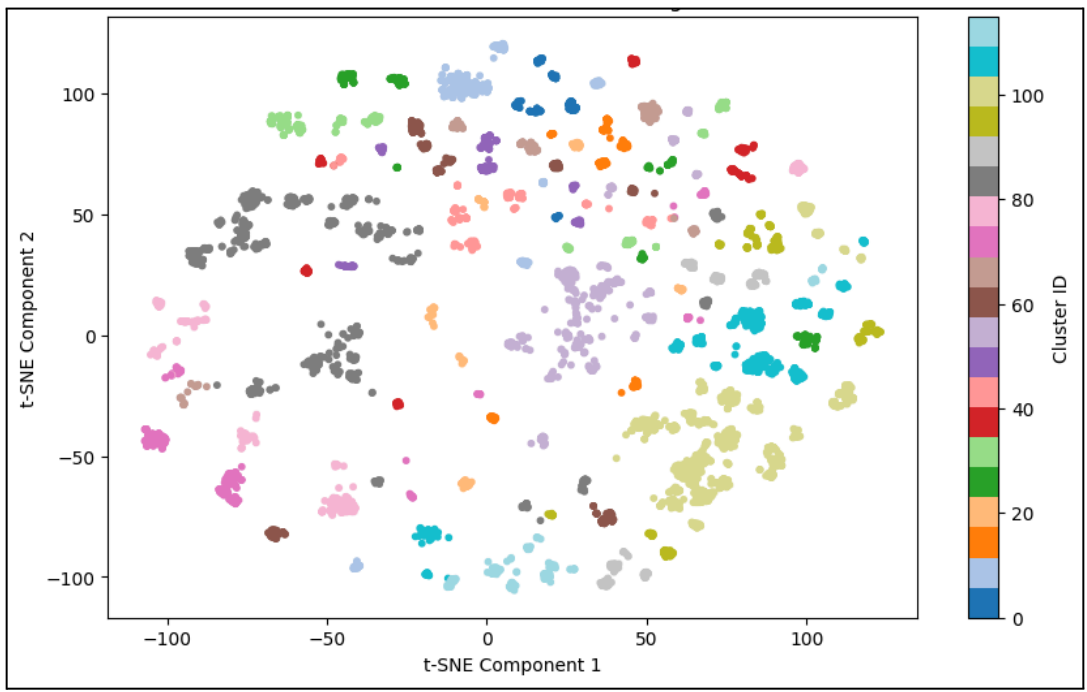

**Cluster Labeling**

To enhance interpretability, each cluster was assigned a meaningful label summarizing its dominant themes. Cluster labels were generated using OpenAI's `gpt-3.5-turbo` model, which processed representative CDEs from each cluster to produce concise and informative names. Specifically, the combined text of up to 20 CDEs per cluster was fed into the model, prompting it to infer key themes and propose suitable labels. The resulting labels reflected distinct groupings in the CDEs, making patterns easier to understand.

For instance, *Cluster 1* was labeled as *"Educational Attainment Levels"*, encompassing 29 CDEs related to education, such as the highest grade completed, degree received, and parental education levels. These CDEs originate from diverse organizations like LOINC, GRDR, NHLBI, and NCI. The automated labeling process effectively captured the common theme, enhancing the interpretability of the cluster. A sample table of CDEs within this cluster is shown in **Table 5**. Supplementary File 1 includes all the clusters along with their corresponding labels.

**Table 5.** Sample of CDEs in Cluster 1 – *"Educational Attainment Levels"*

| tinyId | stewardOrg | Designation | Cluster Name (From LLM) |
|---|---|---|---|
| Co2d1RyYS3 | Project 5 (C…) | Educational Attainment | "Educational Attain…" |
| E8BIucWzGm | ScHARe | Education | "Educational Attain…" |
| XybRLqj3og | LOINC | How far in school did she go Mother | "Educational Attain…" |
| mknGEX8g7 | NEI | Education Level | "Educational Attain…" |
| kxRtCXGZFkc | GRDR | Educational Attainment | "Educational Attain…" |
| 7kBl31ILhQM | NLM | Highest level of education Mother | "Educational Attain…" |
| GESd_nZZSh | NHLBI | How far in school did he go? | "Educational Attain…" |
| P4QmoW1QRq | NICHD | Paternal education | "Educational Attain…" |
| 7y8x7nHXkyl | NCI | Person education level summary type | "Educational Attain…" |
| UGQNVNjCKJ | RADx-UP | What is the highest level of education | "Educational Attain…" |

## Classification

The classification task aimed to evaluate the discriminative power of the generated embeddings for distinguishing between clusters. For each data point, a 1536-dimensional embedding vector was extracted, and used to form the feature matrix. The corresponding cluster labels were utilized as the target variable.

A Random Forest Classifier was selected for this task due to its effectiveness in handling high-dimensional data, its robustness against overfitting, and its interpretability. The classifier, configured with 100 estimators and a fixed random seed, was trained on the training set and subsequently used to predict the cluster labels for the test set. The dataset was split into an 80% training set and a 20% testing set to evaluate model generalization. Model performance was assessed primarily through accuracy[32], and F1-scores[33] across the clusters. The results demonstrated that the embeddings effectively captured cluster-specific characteristics, showcasing their potential for distinguishing between distinct clusters. The high classification accuracy indicates that these embeddings can serve as a strong foundation for downstream AI-driven applications, such as automated annotation, data harmonization, and further refinement with alternative classifiers. While not strictly necessary, a classifier acts as a practical extension of the embedding space, turning its semantic structure into something predictive, interpretable, and efficient. Rather than relying on manual distance calculations, the classifier learns to assign CDE cluster labels quickly, handles complex boundaries, and offers insight into which features matter most. It makes the embedding space operational for real-world use.

# Evaluation

### Clustering Evaluation Using SDOH[34]-Anchored Ground Truth from Gravity Projects[35]

To further evaluate the performance and consistency of the clustering process, we applied the same methodology to a dataset curated by Gravity Projects, focusing on Social Determinants of Health (SDOH) domains. The Gravity Projects dataset groups Common Data Elements (CDEs) into 21 distinct SDOH domains[36], which serve as a benchmark for assessing the quality and relevance of the generated clusters.

This section evaluates the quality of the unsupervised clustering by comparing the resulting clusters against known domain categories, specifically Social Determinants of Health (SDOH) as defined in the Gravity Project. These external references serve as a proxy ground truth to assess cluster coherence and alignment with established healthcare concepts.

The preprocessing steps, including data transformation and embedding generation, were performed in accordance with the process described in the methodology section. This ensured a uniform approach across datasets, allowing for a direct comparison between the results from the NIH NLM CDEs and the Gravity Projects SDOH domains.

## Data Preprocessing and Embedding Generation for SDOH

For the Gravity Projects SDOH dataset, the data includes question-answer pairs related to various social determinants. The specific columns of interest, **Question Concept** (from the screening tool) and **Answer Concept** (from the screening tool), were extracted from the dataset and preprocessed. Any missing values in these columns were replaced with empty strings to ensure consistent input for the embedding process.

**Table 6.** Sample of the final structure with SDOH CDEs to illustrate how the data is organized.

| SDOH Domain Name | Screening Tool Name | Question Concept (from the screening tool) | Answer Concept (from screening tool) |
|---|---|---|---|
| Financial Insecurity | Consumer Financial Protection Bureau (CFPB) | 4.I have money left over at the end of the month | 4- Always; 3- Often; 2- Sometimes; 1- Rarely; 0 -Never |
| Food Insecurity | Medicare THA | Do you always have enough money to buy the food you need | Yes; No |
| Housing Instability | PRAPARE | Are you worried about losing your housing? | Yes; No; I choose not to answer |

To generate embeddings, the **Question Concept** and **Answer Concept** were combined into a single text string for each row. This combined text captured the relationship between the questions and their corresponding answers, which is critical for accurately embedding the data. The combined string was then passed through the OpenAI text-embedding-3-small model to generate embeddings.

These embeddings, now representing the question-answer pairs in a high-dimensional vector space, served as input for the clustering algorithm.

## Evaluation

The primary evaluation criterion was the number of clusters formed after applying the **HDBSCAN clustering algorithm** to the embeddings generated from the Gravity Projects SDOH CDEs. The clustering performance was assessed using several internal validation metrics, including **Silhouette Score**, **Dunn Index**, **Davies-Bouldin Index**, and **Number of Clusters and Outliers** (described in the methodology section). Further external validation metrics[30] were applied using **Adjusted Rand Index (ARI)**[30] and **Normalized Mutual Information (NMI)**[30], as we have the ground truth with the Gravity Projects SDOH domains dataset. External validation metrics are used to assess the quality of clustering by comparing the clustering results to an external benchmark or ground truth. Unlike internal validation metrics

(which evaluate the clustering performance based on the data itself, such as compactness and separation of clusters), external validation metrics require a pre-labeled dataset or ground truth for comparison.

- **Adjusted Rand Index (ARI)**: The ARI values ranged from 0.0062 to 0.6176. ARI measures the similarity between the clustering results and the ground truth, adjusting for chance. A perfect ARI score is 1, indicating complete agreement with the ground truth. The values observed here suggest good agreement with the true SDOH domains, though there is still room for improvement.
- **Normalized Mutual Information (NMI)**: The NMI scores ranged from 0.0604 to 0.8009 **0.71 to 0.81**, indicating a high degree of shared information between the clustering results and the true SDOH domains. NMI values close to 1 indicate a strong correlation between the clustering and the ground truth, and these results reflect strong alignment.

**Table 7.** The results of the evaluation, including various internal and external validation metrics, are as follows:

| min_cluster_size | Silhouette Score | Dunn Index | Davies-Bouldin Index | Number of Clusters | Number of Outliers | ARI | NMI |
|---|---|---|---|---|---|---|---|
| 3 | 0.2325 | 0.3585 | 1.3962 | 82 | 475 | 0.3082 | 0.715 |
| 5 | 0.233 | 0.4215 | 1.4862 | 42 | 612 | 0.4001 | 0.745 |
| 7 | 0.2308 | 0.5123 | 1.5817 | 22 | 722 | 0.5239 | 0.7768 |
| 10 | 0.2067 | 0.4573 | 1.7014 | 16 | 753 | 0.6176 | 0.8009 |
| 15 | 0.0921 | 0.5334 | 1.8472 | 3 | 244 | 0.0136 | 0.0955 |
| 20 | 0.0933 | 0.5345 | 1.8381 | 3 | 273 | 0.0125 | 0.089 |
| 25 | 0.0914 | 0.5358 | 1.8611 | 2 | 292 | 0.0062 | 0.0604 |
| 50 | 0.1936 | 0.5695 | 1.6295 | 2 | 1083 | 0.1978 | 0.394 |

A minimum cluster size of **7** was selected for the final analysis, as it provided a balanced result with a **Silhouette Score of 0.23**, **Dunn Index of 0.51**, and **722 outliers**. This size was deemed optimal for balancing the quality of clusters with the number of outliers. Notably, at a minimum cluster size of 7, the algorithm identified **22 clusters**, which is slightly higher than the **21 ground truth clusters** in the Gravity Projects SDOH domains dataset. This suggests that while the clustering results were largely consistent with the expected structure, the algorithm slightly overestimated the number of distinct clusters. However, this difference is minimal and does not significantly impact the overall clustering performance.

## Confusion Matrix[37] and Clustering Performance

To further assess the alignment between the clustering results and the SDOH domains, we generated a confusion matrix to compare the SDOH Domain names with the cluster names. The confusion matrix allows us to visualize the distribution of data points across the generated clusters and the actual SDOH domains.

- The confusion matrix for the SDOH Domain Name vs. Cluster Name highlights how well the clustering algorithm mapped the SDOH domains to the generated clusters.
- The final analysis showed that the majority of the SDOH domains were well-represented within the generated clusters, with only a few discrepancies that could be explored further.

**Figure 4.** Confusion Matrix of SDOH Domain Names vs. Generated Clusters

Note: "Outliers" were excluded from the confusion matrix to avoid distortion of the results. These points, labeled −1 by HDBSCAN, do not belong to any cluster and including them could introduce noise.

## Classification Evaluation Using LLM-Derived Labels from NIH CDEs

In contrast to the clustering evaluation, this classification task does not rely on external ground truth. Instead, the cluster labels—automatically generated by a LLM via summarization of NIH CDEs—serve as the target variable. The goal is to assess the discriminative power of the generated embeddings by training a classifier to predict these LLM-derived labels. As detailed in the Methodology section, a Random Forest classifier was trained on the embedding vectors, and its performance was evaluated using standard classification metrics.

The classification model achieved an overall accuracy of 0.9046, demonstrating strong performance across several categories. However, a detailed analysis of the precision, recall, and F1-score for each class reveals noteworthy variations in performance (see Supplementary File 2 for the complete classification report[38]).

The model performed exceptionally well in categories with larger sample sizes, such as the "Mail Address Demographics Cluster," where precision and recall reached 0.89 and 1.00, respectively, yielding an F1-score of 0.94. Other high-performing categories include the "Cognitive Assessment and Medical History Cluster (CAMHC)" with an F1-score of 0.98 and the "Cognitive Function and Behavior Assessment Cluster," which also achieved an F1-score of 0.98.

Conversely, the model struggled with classes that had fewer samples or exhibited greater complexity, such as the "Life Challenges and Personal Struggles" and "Amygdala-Neuronal-Loss-Severity Cluster," where both precision and recall were 0.00. Additionally, several clusters with smaller support values experienced similar challenges, likely due to data imbalance and insufficient training examples.

The precision-recall tradeoff in certain categories highlights the need for further refinement. For instance, while the "High-Risk Drinking Patterns Scale" attained a balanced precision and recall of 0.94, resulting in a high F1-score of 0.94. In contrast, the "Hand and Body Tremor and Functionality Scale" exhibited a stark contrast with a precision of 1.00 but a recall of only 0.11, resulting in a lower F1-score of 0.20.

Future efforts should focus on strategies to mitigate data imbalance, such as oversampling underrepresented classes or employing class-weight adjustments during training. Additionally, incorporating techniques like ensemble learning or fine-tuning hyperparameters may enhance performance consistency across all categories.

Overall, the model shows promise, particularly in well-represented clusters, while presenting opportunities for improvement in handling sparse and complex categories.

# Conclusion

This research introduces a dynamic and scalable framework to facilitate the harmonization of Common Data Elements (CDEs) across heterogeneous biomedical datasets using LLM embeddings and unsupervised clustering techniques. Our approach addresses key challenges in data harmonization: semantic heterogeneity, structural variability, and context-dependence. By leveraging OpenAI's text-embedding-3-small model and HDBSCAN clustering with an optimized minimum cluster size of 20, we identified 118 meaningful clusters in the NIH NLM CDE Repository. The classification model achieved 90.46% accuracy overall, with particularly strong performance in well-represented categories.

External validation against Gravity Projects' Social Determinants of Health domains demonstrated strong alignment, with an Adjusted Rand Index of 0.52 and Normalized Mutual Information of 0.78. These results confirm that our generated embeddings effectively capture cluster-specific characteristics.

A notable challenge observed in our analysis was the high proportion of outliers – 17,973 out of 24,363 CDEs (73.8%) in the NIH NLM dataset and 722 out of 1,335 (54.1%) in the Gravity Projects dataset. This suggests significant heterogeneity in the data that could benefit from more nuanced clustering approaches or domain-specific pre-processing. Future work should investigate these outliers to determine whether they represent truly unique elements or potential subclusters that could be meaningfully grouped with refined techniques.

Another limitation stems from the fixed input size for cluster labeling—specifically, limiting each cluster to 20 CDEs due to the 4,096-token constraint of GPT-3.5-Turbo. While this ensured compatibility and efficiency, it may oversimplify larger clusters with diverse content, limiting the semantic richness of the generated labels. This also presents room for future optimization through more adaptive input selection or dynamic labeling strategies.

Our framework is designed with potential for integration with existing CDE repositories like caDSR, CEDAR, and METEOR. This compatibility would allow research teams to adopt the framework as a complementary tool alongside their current data infrastructure, promoting wider adoption and maximizing utility across the biomedical research ecosystem.

While the framework performs exceptionally well for larger clusters, we observed limitations in handling underrepresented categories. Future work should focus on addressing data imbalance through techniques such as oversampling or class-weight adjustments during model training. Additionally, exploring alternative classifiers like Support Vector Machines, XGBoost, and Neural Networks may further enhance performance under different conditions.

As biomedical datasets continue to grow in size and complexity, implementing vector databases would significantly improve scalability and query performance. Vector databases specialized in similarity searches could accelerate retrieval operations and support real-time harmonization of new CDEs, making this framework more viable for production environments with continuously expanding datasets. While our current implementation employs dimensionality reduction techniques that theoretically support scaling, further performance testing and optimization will be necessary as datasets grow to millions of entries.

Our framework offers a flexible solution to the persistent challenge of harmonizing Common Data Elements (CDEs). By converting isolated datasets into AI-ready formats with minimal manual intervention, it facilitates more effective data integration and reuse.

## Supplementary Materials

The following supplementary files are available to support the findings and analysis described in this manuscript:

- **Supplementary File 1**:
  NIH_NLM_CDE_Clusters.xlsx This file contains the final set of CDE clusters generated by the framework, along with their assigned labels.

- **Supplementary File 2**:
  Classification_report.txt This text file includes detailed classification metrics (precision, recall, and F1-score) for each class label used to evaluate the quality of the cluster assignments.

- **Supplementary Code**:
  CDE_methodology.ipynb A Jupyter Notebook containing the implementation of the methods described in the manuscript, including data preprocessing, embedding generation, clustering, and label assignment. This is provided to support reproducibility and transparency of the methodology.

## Acknowledgments

Funding provided by
- NIH NHLBI INCLUDE 5U2CHL156291
- BDC DMC 1OT2HL167310

Support for this work was provided by the National Institutes of Health, National Heart, Lung, and Blood Institute, through the BioData Catalyst program (award 1OT3HL142479-01, 1OT3HL142478-01, 1OT3HL142481-01, 1OT3HL142480-01, 1OT3HL147154-01). CJM was funded in part by the Director, Office of Science, Office of Basic Energy Sciences, of the US Department of Energy DE-AC0205CH11231. Any opinions expressed in this document are those of the author(s) and do not necessarily reflect the views of NHLBI, individual BioData Catalyst team members, or affiliated organizations and institutions.